\documentclass[10pt,letterpaper]{article}
\usepackage[top=0.85in,left=2.75in,footskip=0.75in]{geometry}
\usepackage[UKenglish]{babel}
\usepackage{amsmath,amssymb}

\usepackage{changepage}

\usepackage{textcomp,marvosym}

\usepackage{cite}

\usepackage{nameref,hyperref}

\usepackage[right]{lineno}

\usepackage[nopatch=eqnum]{microtype}
\DisableLigatures[f]{encoding = *, family = * }

\usepackage[table]{xcolor}

\usepackage{array}

\newcolumntype{+}{!{\vrule width 2pt}}

\newlength\savedwidth

\raggedright
\usepackage[aboveskip=1pt,labelfont=bf,labelsep=period,justification=raggedright,singlelinecheck=off]{caption}

\makeatletter
\renewcommand{\@biblabel}[1]{\quad#1.}
\makeatother

\usepackage{lastpage,fancyhdr,graphicx}
\usepackage{epstopdf}
\fancyheadoffset[L]{2.25in}
\fancyfootoffset[L]{2.25in}
\begin{document}
\vspace*{0.2in}

\begin{flushleft}
{\Large
\textbf\newline{Directionality measures in evolutionary ecological networks: Insights from the Tangled Nature model} 
}
\newline
\\
Andrea Marchetti\textsuperscript{1,2,3},
Henrik Jeldtoft Jensen\textsuperscript{3*}
\\
\bigskip
\textbf{1} Department of Physics and Astronomy, University of Padua, Padua, Italy
\\
\textbf{2} Galilean School of Higher Education, University of Padua, Padua, Italy
\\
\textbf{3} Centre for Complexity Science and Department of Mathematics, Imperial College London, London, United Kingdom
\\
\bigskip

%
%





* Corresponding author \\ Email: h.jensen@imperial.ac.uk (HJ)

\end{flushleft}
\section*{Abstract}
The myriad microscopic interactions among the individual organisms that constitute an ecological system collectively give rise, at the macroscopic scale, to evolutionary trends. The ability to detect the directionality of such trends is crucial for understanding and managing the dynamics of natural systems. Nevertheless, identifying the key observable quantities that capture such directional behaviour poses a major challenge. In this study, we propose that translating ecological data into a network framework is a valuable strategy to measure system stability and evolution. We examine the Tangled Nature model as a test case, evaluating network entropy, species diversity, and the clustering coefficient as metrics of network stability and directionality.



\section*{Introduction}

\par Before we turn to evolutionary ecology, let us recapture how the direction of macroscopic processes is handled in thermodynamic systems. Namely, the macroscopic behaviour emerges from the collective dynamics of numerous microscopic phenomena. These involve the movements and energy exchanges of particles at the atomic or molecular level, each following its own path governed by fundamental physical laws. Understanding how these microscopic phenomena aggregate to produce macroscopic behaviour presents a significant challenge due to the very high number of variables and the intricate and often non-linear nature of their interactions.
Thermodynamics, however, provides a robust framework to navigate these complexities. By employing a few scalar macroscopic quantities like temperature, pressure, and energy, it allows us to make accurate predictions about the overall behaviour of a system. These macroscopic quantities encapsulate the collective effects of countless microscopic interactions, thereby offering a practical and manageable way to understand and predict certain systemic-level properties of the behaviour of many interacting components. 

\par A key role in this is played by entropy, understood as a measure of the disorder in a system, reflecting the number of possible microscopic configurations. By looking at a single scalar quantity that summarises the microscopic dynamics, we can in specific cases make powerful predictions about the direction of the system's overall evolution. The second law of thermodynamics states that for a closed system the entropy cannot decrease as a consequence of the unfolding of internal processes.  In this sense, the directionality of the physical evolution of the system is determined.

\par Of course living systems are not closed systems, but they do consist of very many components. At various levels, biological systems are shaped by countless microscopic interactions that collectively determine their macroscopic behaviour. Cells and their functioning arise from the combination of biological and non-biological molecules, their transport phenomena, and their chemical reactions. Multicellular  organisms are the result of the myriad interactions between billions of cells of different types. Ecological systems emerge from the interplay of many different individuals in a population and of populations of different species. 

\par In particular, evolutionary processes consist of continuous directional and adaptive changes in the composition of types present in an ecosystem, as one population type replaces another over time due to mutations, competition, and selection. At the microscopic level, these processes are characterised by pairwise species competition between an established type and a mutant, and the outcome consists in either the extinction or the fixation of the mutant. At the macroscopic level, the combination of many substitution processes leads to changes in the composition of ecological communities over time, with some new species developing from ancestral types and some disappearing because of endogenous (competition with new, more suitable species) or exogenous factors (changes in the environmental conditions).

\par This multitude of processes and effects reminds one of statistical thermodynamics and suggests that macroscopic information might be extracted from a few scalar quantities. Can we define a quantity for living systems that shows a directional behaviour in a way similar to entropy in thermodynamics? Can we systematically predict the outcome of species competition? Can we tell which set of species is favoured as we can tell which configuration of a thermodynamic system is favoured?

\par 
Crucially, living organisms are open systems and out of equilibrium, in sharp contrast to the systems studied by classical thermodynamics. They display the ability to maintain and create order despite the entropic tendencies of their surroundings, but they achieve this by continuously importing energy and matter from their environment and exporting entropy and waste.  Consequently, statistical equilibrium thermodynamics is not directly appropriate for the study of the dynamics of these systems.

\par An effective study of directionality in evolutionary ecology must therefore consider relevant quantities that fall outside the scope of classical thermodynamics. The challenge is now to 
determine the observables that are determinant for the outcome of the competition between the coexisting types. When neglecting stochastic effects -- which might lead a small population to extinction independently of its success potential -- various theories have been proposed. According to r/K selection theory, first proposed by McArthur and Wilson \cite{MacArthur2001}, organisms might prefer offspring quantity or offspring quality depending on the environmental conditions. In unstable environments, r-selected species might have an advantage: the determinant factor for evolutionary success is the growth rate r. These species favour a fast reproduction to outgrow competing species. On the other hand, in slowly changing environments K-selected species can dominate: when living at densities close near carrying capacity K, the determinant factor is how competitively a species collects shared resources. Life History Theory has then incorporated and developed the r/K-selection paradigm while stressing the importance of the life-cycle of each organism \cite{Pianka1970, Reznick2002}. In general, however, species are rarely strictly r- or K-selected; instead, they are found along an r/K continuum.

\par These theories assume that two main factors determine the outcome of the competition: the offspring production rate and the resource uptake rate. Some species might be favoured by their ability to produce more offspring in a given amount of time, others by the fact that they collect more resources because of a more complex metabolism, leaving less resources available to the rest of the ecological community. Although these two selection mechanisms jointly play a role in the evolution of real ecological systems, classical models of evolution consider only one factor at a time. 

\par In bioenergetic models, rooted in Lotka's studies \cite{Lotka1922}, the evolutionary advantage of one type over another is determined by thermodynamic entropy \cite{Demetrius2013}. This assumes that entropy production can measure a system's efficiency in extracting and distributing energy. The rate of matter and energy circulation reflects this rate of entropy production. As more efficient types replace less efficient ones, both entropy production and energy circulation increase. However, these models assume negligible resource input, which limits their validity in open living systems. On the other hand, demographic models, starting with the work of Fisher \cite{Fisher1930}, individuate the selective advantage as the growth rate, but neglect the interaction with the environment, and only work in the limit of infinite resources.

\par In the 1970s, based on Leslie models \cite{Leslie1945}, Lloyd Demetrius introduced \textit{Evolutionary Entropy} to incorporate the two mentioned selection mechanisms \cite{Demetrius1974, Demetrius1974isomorphism, Demetrius1975, Demetrius1975selection}. Most significantly, mathematically, Evolutionary Entropy obeys a powerful directionality principle analogous to the second law of thermodynamics, providing insights into the outcome of species competition based on the type of resources available and the structure of the organism's life cycle. 

 Interestingly, the directionality depends on the specific conditions of the surrounding environment. Systems with a constant level of diverse resources will favour organisms with higher Evolutionary Entropy, while systems with resources with variable abundance and singular composition will favour organisms with lower Evolutionary Entropy \cite{Demetrius2013}. In any case, Evolutionary Entropy can be used as a single scalar indicator to suggest the most probable outcome of species competition. 
 

\par Despite the significant insights given by the directionality principle, Evolutionary Entropy is defined only on a very specific class of models. While the Leslie model effectively captures the details of individual species' life cycles under constant conditions, it is limited in its capacity to address broader aspects of community evolution. For instance, it lacks the ability to naturally incorporate essential factors such as time-varying birth and death rates and environmental variability \cite{Caswell2001}, as well as non-competitive interactions, the presence of a high number of species, and mutations.


\par To overcome these limitations, a more comprehensive evaluation of directionality in evolutionary ecology should include such factors. In particular, agent-based models can be useful tools to study how macroscopic patterns emerge from simple microscopic dynamical rules.


\par  In this work, we focus on the Tangled Nature  model \cite{Christensen2002, Jensen2018}, a simple agent-based computational model of co-evolutionary ecology that naturally reproduces complex systemic-level ecological phenomena such as the structure of co-existing species networks, the intermittent replacement of ecological communities through mass extinctions, as well as ecological observables such as Species Abundance Distributions (SADs) and Species Area Relations (SARs).  Most notably, the TaNa model exhibits a clear directionality emerging from its non-directional microscopic dynamics. As the system evolves, the robustness of the ecological communities against stochastic fluctuations increases, driven exclusively by the model's internal dynamics. Hence, the model’s balance between simplicity and its accurate replication of real-life ecological phenomena makes it an appropriate proxy for testing directionality measures that could apply to real ecological systems.



In this paper, we rephrase the TaNa in terms of a network problem and study its directionality by means of network measures. The rest of the paper is organised as follows. First, we consider network entropy, defined as an extension of Evolutionary Entropy on networks \cite{Demetrius2005}, as a potential indicator of the robustness of ecological communities, searching for any directional trend indicating an increase in stability over time. We find that, when analysing the average behaviour across a set of simulations, a clear increasing trend in network entropy is observed, which confirms the overall increase in stability of TaNa ecological communities over time. We then demonstrate that network entropy is in fact directly connected to TaNa ecosystems' diversity, and thus propose a validation of the historically well-know diversity--stability relation.


Next, to corroborate the results obtained, we complement our study by considering additional network stability measures. We identify the clustering coefficient as an effective metric for this purpose, as it provides an approximate measure of the network's local interconnectedness. In particular, we separate the network into two parts, one consisting of positively (mutualistic) interacting species and another consisting of negatively (antagonistic) interacting species. Our simulations show that the clustering coefficient remains stable for the positively interacting subnetwork and decreases for the negatively interacting one, suggesting a weakening of the parasitic portion of TaNa ecological communities over time and hence an overall increase in stability. Lastly, we examine in greater detail the specific interaction structure of each ecological community, finding that mutualism becomes predominant as the system evolves, confirming a trend towards greater stability in TaNa simulations.

\section*{The Tangled Nature model}

\par The Tangled Nature (TaNa) model \cite{Arthur2017, Lawson2006, Anderson2005, Becker2014, Laird2006, Jensen2018, Hall2002, Christensen2002} is a simple agent-based model of ecological co-evolution in which population dynamics includes species interaction and mutation. In this model, species, or types, are individuated by a unique label. The likelihood of individual reproduction is determined by the interaction with other types and resource limitation is represented by a term similar to a carrying capacity. In contrast, mutation and death are assumed to occur randomly with equal probability for all types.

Despite its simplicity, and far from being a comprehensive representation of a real ecological system, the model is able to emulate some of the main characteristics of
the long-term dynamical behaviour of evolutionary processes. Above all, it exhibits segregation in the type space at the macroscopic level, corresponding to speciation, and intermittent behaviour, consisting in a succession of metastable ecological communities separated by transition periods that correspond to mass extinctions. This intermittency is reminiscent of punctuated equilibrium \cite{Gould2002} and interestingly induces a precise directionality that arises as an emergent property from the microscopic dynamics at the agent level. In particular, the lifetime of the metastable communities increases \cite{Christensen2002,Hall2002} as internal evolutionary selection succeed in generating sets of types whose interaction network is more robust, typically  more mutualistic \cite{DiazRuelas2016}.

\subsection*{Definition of the model}

\par In the TaNa model, a type $a$ is defined by a simple label represented by a sequence of binary values with fixed length $L$, denoted $\mathbf{S}^a = (S^a_1,\dots,S^a_L)$ with $S_i^a = \pm 1$. 
The sequence $\mathbf{S}^a $ is not intended to have any direct physical meaning; the similarity to a genome is obviously very crude. We use this format for the labels of the different types because it makes representation of the effect of mutations simple.


Each sequence $\textbf{S}^a$ represents one position in  \emph{type space} $\mathcal{S}$. This space is represented geometrically by an L-dimensional hypercube $\mathcal{S} = \{-1, 1\}^L$ of size $2^L$. Each agent will belong to a specific position $\textbf{S}^a$ and the number of individuals of type $a$ at time $t$, i.e. the \emph{occupancy}, is denoted $n(\textbf{S}^a, t)$, while the total population at time $t$ is given by $N(t) = \sum_{a=1}^{2^L} n(\textbf{S}^a,t)$.

\par The system's dynamics is studied by  use of Monte Carlo 
simulations. A time step consists of one attempt of offspring production, including possible mutation, and one killing attempt. During the latter, an individual is chosen randomly from the population and killed with a probability $p_{kill}$ constant in time and independent of the type. We let the chance to be killed be the same for all types, since death often occurs accidentally independent of a type's abilities. Moreover, Darwinian evolution emphasises the importance of the relative reproduction rate as a driver for selection amongst different types. 

\par After the killing attempt, a different randomly chosen individual of type ${\bf S}^a$ is chosen for reproduction. For simplicity we consider asexual reproduction: the parent produces two offspring of the same type. The parent is then removed from the system. The reproduction happens with probability  $p_{off}(\textbf{S}^a,t) = \frac{\exp[\mathcal{H}_W(\textbf{S}^a,t)]}{1+\exp[\mathcal{H}_W(\textbf{S}^a,t)]}$, which depends on the interaction with all existing types at time $t$ via the weight function
\begin{eqnarray}
\label{eq:Hamiltonian}
    \mathcal{H}_W(\textbf{S}^a,t) = \dfrac{K}{N(t)} \sum_{\{\textbf{S}^b\}} J(\textbf{S}^a,\textbf{S}^b)n(\textbf{S}^b,t) - \mu N(t).
\end{eqnarray}
In Eq \eqref{eq:Hamiltonian}, the interaction strength $J(\textbf{S}^a,\textbf{S}^b)$ couples species $a$ and $b$, while $\mu$ corresponds to the inverse of a carrying capacity. Consequently, positive interactions with other species will increase the probability of reproducing, while negative interactions and a high number of individuals $N(t)$ in the system will reduce it. 

\par After birth, each of the offspring may mutate. This is represented by allowing each of the ``genes" $S^a_i $ to mutate independently, that is, $S^a_i \mapsto -S^a_i$ with probability $p_{mut}$. 

\par As always in Monte Carlo simulations, one combines a number of time steps into a macroscopic physical time independent of the size of the system. We do this by defining  a \textit{generation} as $N(t)/p_{kill}$ time steps. This corresponds to the average time necessary to kill all individuals present in the system at time $t$.

\subsection*{Interaction matrix}
\par The interaction matrix $J$ in the weight function \eqref{eq:Hamiltonian} represents the influence of one possible type on another. In principle, these influences could be measured experimentally. For example, one could study the growth rates of two types of bacteria -- first in isolation and then when co-evolving on the same Petri dish. However, in practice, the exact values of the weights $J(\textbf{S}^a,\textbf{S}^b)$ are unknown. Therefore, the matrix $J$ is generated randomly \cite{Christensen2002, Jensen2018, Hall2002}. This means that its entries are selected randomly at the start of the simulation and remain fixed thereafter.


Different assumptions about the statistics of $J$ influence the emergent properties of the generated communities of types \cite{Anderson2005}. However, the behaviour essentially consists of only two modes. We denote by $\theta$ the fraction  of all possible pairs that interact. This means that $\theta$ represents the connectivity of $J$ when considered as an adjacency matrix. For very low values of $\theta$, the communities of interacting types split into disconnected subcommunities. The SADs for such low values of $\theta$ are very different from the approximately log-normal forms that are often observed in nature.  For $\theta$ values high enough to produce communities that form a single connected subnetwork of interacting types, the SAD is compatible with ecological observations and the dynamics consistently exhibit the intermittent appearance of metastable communities for broad ranges of the parameters.


\par In this work we use a version of the model that excludes self-interactions, i.e. $J(\textbf{S}^a,\textbf{S}^a) = 0~\forall a$. While including self-interactions does not imply any changes in the qualitative behaviour of the model, the hypothesis of null self-interaction highlights the effect of co-evolution and prevents species from living and reproducing in isolation.

\par On the other hand, as already mentioned, the interaction strengths  $J(\textbf{S}^a, \textbf{S}^b)$ and $J(\textbf{S}^b, \textbf{S}^a)$ are fixed at values drawn independently from a given probability distribution\cite{Christensen2002,Hall2002,Jensen2018}. The resulting interaction matrix is therefore  asymmetric. The behaviour of the model depends on the maximum permitted interaction strengths. When interactions are weak, no intermittency is observed. However, when strong interactions occur with a non-zero probability, the system exhibits intermittent macro-dynamics.

The distribution of interaction strengths used typically consists of a symmetric distribution on the interval $[-K,K]$ peaked at the origin  \cite{Christensen2002,Hall2002,Jensen2018} or a Gaussian  \cite{Laird2006}. Since the interaction matrix is asymmetric, we can interpret the relationship between type $\textbf{S}^a$ and $\textbf{S}^b$ in  the following way: mutualistic ($J_{ab} > 0$ and $J_{ba} > 0$), antagonistic ($J_{ab} < 0$ and $J_{ba} < 0$), and predator-prey ($J_{ab} < 0$ and $J_{ba} > 0$ or vice versa). In a large enough matrix, all possibilities will be present in a significant proportion.

\subsection*{Dynamical Behaviour}

\par
\begin{figure}[!h]
\includegraphics[width=\textwidth]{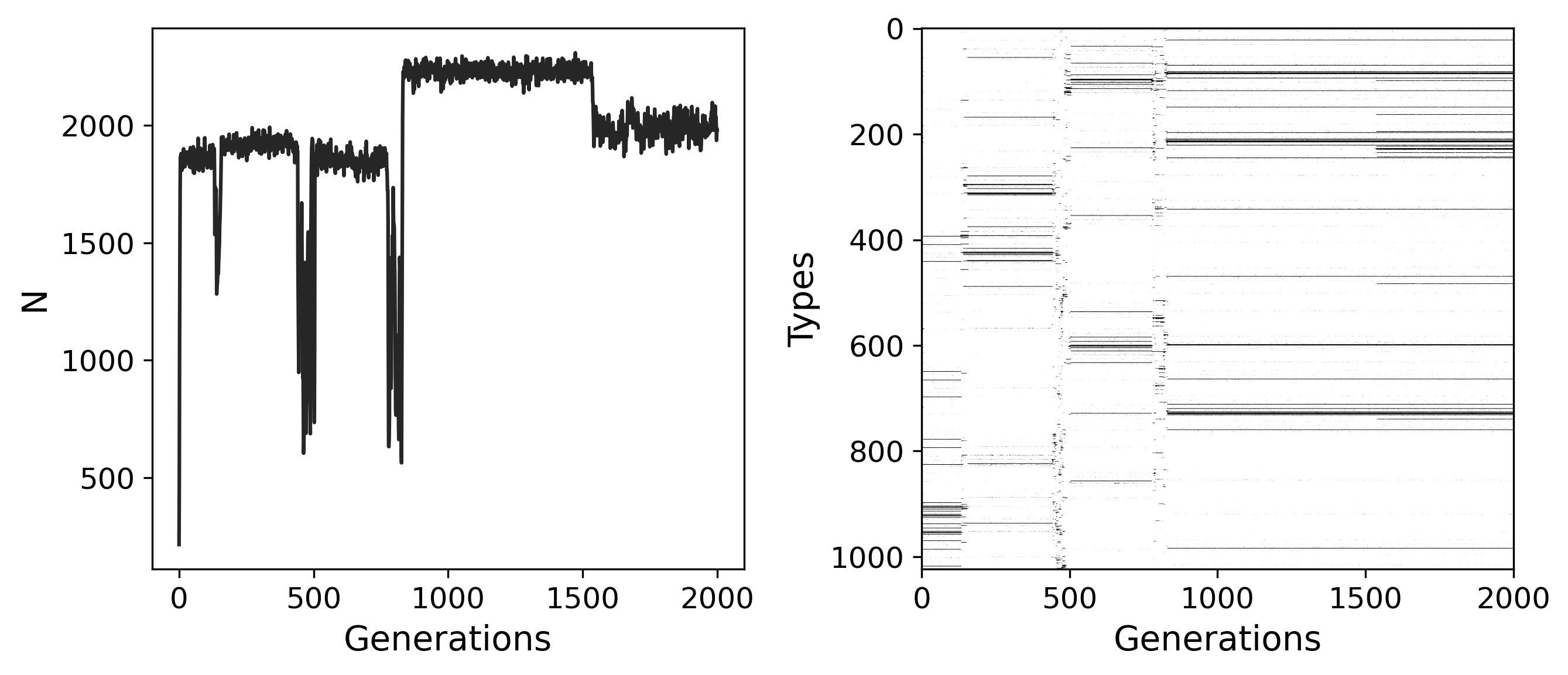}
\caption{{\bf TaNa simulation.}
Left: Total number of individuals as a function of time (in generations) for a single realisation of the TaNa model with parameters $L = 20$, $\mu = 0.1$, $K = 0.01$, $p_{kill} = 0.2$, $p_{mut} = 0.01$. Right: Occupancy distribution of the types in the same simulation. The types are labelled arbitrarily and a black dot indicates a type occupied at the time $t$. The punctuated dynamics is clearly visible: long metastable periods alternate with short periods of chaotic transition, in which $N(t)$ exhibits large amplitude fluctuations.}
\label{fig:TaNa}
\end{figure}

 The microscopic rules described generate intermittent macro-dynamics characterised by metastable states called quasi-Evolutionary Stable Strategies (qESSs). These states remain stable for a certain duration, though mutations cause fluctuations in occupancy during a qESS state (see Fig~\ref{fig:TaNa}). As mutants continuously explore the viability of nearby configurations, a transition to a new qESS state is eventually triggered. These transitions involve adaptive random walks through configurations in type space while searching for a new metastable state and are associated with high-amplitude fluctuations of $N(t)$.

\par During a qESS state, the system contains a limited number of persistent, mutually supportive types. These types show high, stable occupancies throughout the whole qESS state and we call this set of types the \emph{core}. Each core type is typically surrounded by a cloud of mutants with low occupancy.  These \textit{cloud} types are typically unstable and will continuously appear and disappear. Therefore, speciation can be thought of as the segregation of highly occupied types from the set of all the possible types in the type space. 

\par On the other hand, during a transition -- which usually lasts for a smaller number of generations -- diversity \cite{Comment_diversity} dramatically increases, with a high number of different types newly appearing, and the system is characterised by a hectic rearrangement of the composition of types in the entire population.

\par We emphasise that the model is agent-based and driven by three stochastic processes: reproduction, mutation, and killing. The stochastic dynamics is responsible for the fluctuations observed during a qESS state and in the transition between successive qESS states. Indeed, the term ``quasi" reflects the fact that stochastic mutations can jeopardise the stability of an ESS state, potentially triggering a transition to a completely different state in the configuration space.

\par Furthermore, one of the main strengths of the TaNa model is the inherently co-evolutionary nature of its dynamics. It does not make use of a fixed fitness landscape, but assumes that the environment of a type is given by the combined effect of coexisting types. Indeed, the evolution of each species is largely determined by the interaction with the other extant species, which is proportional to their occupancies. Since the number of individuals of each species constantly varies throughout the system's evolution, the fitness landscape is continuously modified as well. Fluctuations in the stochastic dynamics may cause the instantaneous environment to change as new mutations emerge, potentially leading to transitions to new qESS configurations.

\subsection*{Long-term behaviour}

\par  Despite the model's simple dynamics, it shows a striking overall agreement with real systems. The SAD and SAR of the TaNa have the same form as those often observed in nature \cite{Lawson2006, Anderson2005}, with the former appearing in a log-normal form and the latter showing the sub-linear increase typical of medium (regional) scales \cite{Rosenzweig1995}. In addition, the TaNa is particularly successful in demonstrating a connection between complexity and stability, as well as adaptive behaviour over time\cite{Jensen2018}. 
\par The average long-term behaviour of the TaNa can be studied by averaging over an ensemble of simulations, each run for a large number of generations. This has been studied extensively in previous work \cite{Christensen2002, Hall2002, rikv03:punc,  Laird2006, Murase2010, sevi05:effe, Becker2014, Sibani2013}, proving that the TaNa shows a clear directional behaviour in that it displays, on average, a general increase in stability over time. This can be deduced from the increase in the duration of qESS states and decrease in the occurrence of transitions, implying a growth in the resistance to perturbations of ecological communities as they evolve \cite{Comment_stability}. In addition, the diversity and average number of links per species also increase so that, as the system evolves, it tends to display a higher number of species that are more interconnected to each other. This can be interpreted as an increase in the complexity of the ecological communities, reproducing what is observed in nature and challenging the complexity--instability paradigm first proposed by Robert May \cite{May1972}. Finally, the dynamics is characterised by an increase in mutualism -- consistent with some experimental observations \cite{Leigh2010} -- which, in turn, enhances the system's stability.

\par These findings collectively demonstrate that the system tends to spontaneously evolve towards more stable and complex ecological communities, in terms of population distributions in the type space, as is often observed in nature \cite{McCann2000}. This suggests that the TaNa model can effectively capture ecological evolution. Indeed, as time passes, the ecological networks become increasingly better adapted: the ecological community as a whole reaches collectively increasingly stable configurations among the total set of all possible ways of distributing a population through type space. 

\par Notice that, in these terms, selection and adaptation operate at the level of the entire configuration in the type space rather than at the level of individual types. This highlights that the biological concept of fitness makes most sense when considered as a collective property of an ecosystem, rather than an inherent characteristic of the individual species or individual members of a population, as it is sometimes traditionally assumed. 

\par In summary, the TaNa model demonstrates a macroscopic directionality that arises purely from the underlying microscopic dynamics. The large-scale patterns and long-term trends within the system are directly and exclusively the result of the interactions and dynamics occurring at the microscopic level, without the need for additional external influences.

\subsection*{The Tangled Nature as a network problem}

\par In this section, we propose an approach in which we simplify the TaNa model by translating it into a network problem, focusing on the networks associated with TaNa's ecological communities. These networks are defined by the effective interaction matrices that drive the TaNa dynamics. This operation allows us to analyse the evolution of robustness in these networks, providing two new proper measures of stability in the TaNa. 
We recall that  existing knowledge about the evolution of stability in the model is derived from interpreting the duration of quasi-stable states and the frequency of transitions, which are not direct indicators of stability themselves.

\par One of the key elements of the TaNa model is the interaction matrix $J_{ij} \equiv J(\textbf{S}^i, \textbf{S}^j)$, which defines the sign and magnitude of the interaction strength between any two types $i$ and $j$ and thereby affecting the offspring probability of the species considered. In particular, the effective interaction matrix defined as 
\begin{eqnarray}
\label{eq:Jn}
    (Jn)_{ij}(t) \equiv J_{ij}n_j(t)
\end{eqnarray}
determines, at each simulation time step $t$, the specific set of active links connecting any extant type $i$ to a coexisting type $j$, that is, $n_i(t),n_j(t)>0$ and  $J_{ij}\neq0$. This quantity is crucial for driving the TaNa dynamics due to its role in the weight function $\mathcal{H}_W$ (see Eq \eqref{eq:Hamiltonian}). 

\par
\begin{figure}[!h]
\centering
\includegraphics[width=0.5\textwidth]{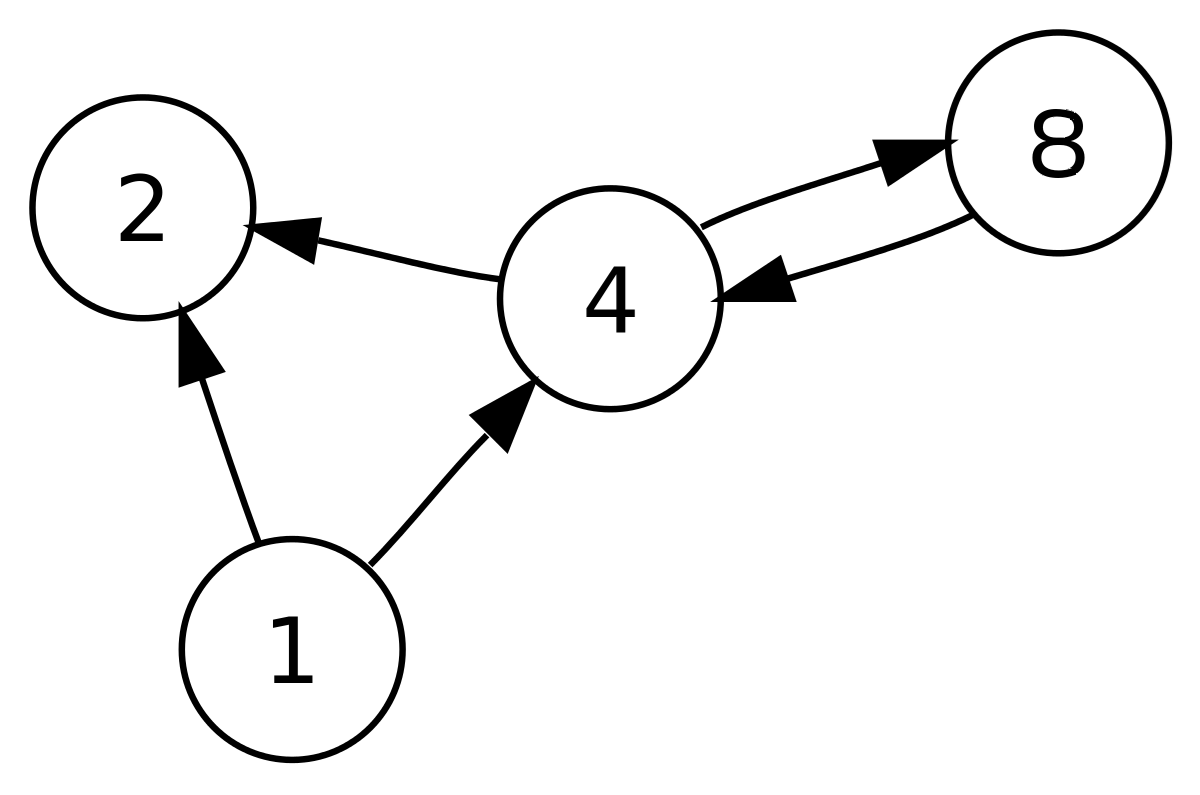}
\caption{{\bf TaNa ecological network.}
Schematic representation of the interaction network representing the ecological community generated by a TaNa simulation at a given time step. Each node represents a type, arbitrarily labelled, so that only the species with at least one individual appear, while edges represent the interaction between two species, weighted as $J_{ij}n_j$ for an edge connecting $j$ to $i$.}
\label{fig:TaNa_network}
\end{figure}

 Hence, we map each ecological community at time $t$ -- as defined by a set of species and their interactions -- onto  a  network with adjacency matrix $Jn$ (Fig \ref{fig:TaNa_network}). This forms the basis on which to compute appropriate network measures that extract information about the stability of the underlying ecological communities.

\section*{Network measures}

\subsection*{Network entropy}

\par Consider an undirected, unweighted graph whose adjacency matrix is given by $A$. Since each node represents a different type, connections between nodes represent a generic flow of energy and matter between them. The Markov matrix $P$ associated to such flow can be computed from $A$ as \cite{Demetrius2005}
\begin{eqnarray}
\label{eq:P_ij}
	P_{ij} = \frac{A_{ij} u_j}{\lambda u_i},
\end{eqnarray}
where $\lambda$ is the dominant eigenvalue of $A$ and $\mathbf{u}$ the associated eigenvector, as long as the Perron-Frobenius theorem holds for $A$. The latter condition is verified for a positive matrix or for a non-negative irreducible matrix. In particular, for an undirected, unweighted graph, which is symmetric and has non-negative entries $A_{ij} = A_{ji} \in \{0,1\}$, the Perron-Frobenius theorem holds as long as the graph is strongly connected.

\par Let $\boldsymbol{\pi}$ denote the stationary distribution of the Markov process described by $P$. The associated network entropy is defined as \cite{Demetrius2005}
\begin{eqnarray}
\label{eq:H}
	H = - \sum_{i,j} \pi_i P_{ij}\log P_{ij}.
\end{eqnarray}
In this context, network entropy can be seen as a measure of the multiplicity of network paths (mathematical details are provided in S1 Appendix). In particular, network entropy will be minimum ($H = 0$) when all the network's nodes are connected by a unique path, and maximum ($H = \log N$ with $N$ the number of nodes) when the network is a complete digraph.

\subsubsection*{Network entropy as a measure of robustness} 

\par Importantly, network entropy can be used as an indicator of  robustness defined in general  as the network's ability to preserve its functionality following node removal \cite{Demetrius2005, Tejedor2017} or the insensitivity of
 network's parameters to dynamical changes in the individual variables \cite{Demetrius2005}. For example, robustness  in ecological networks  corresponds to the survival capacity of extant species as invasive ones are added to the network, or as native ones are removed from it \cite{Tejedor2017}. In particular, network robustness appears to be strongly related to loopiness \cite{Kwon2007, Chujyo2021} and connectivity. Network entropy, being related to the number of possible paths in the network, constitutes a measure of network connectivity and  loopiness, and can therefore be used as an appropriate indicator of network robustness \cite{Demetrius2005}.

 \par In particular, a result known as the Fluctuation-Stability Theorem (FST) \cite{Demetrius2013} states that network entropy is positively correlated with the fluctuation decay rate for a sufficiently small perturbation magnitude \cite{Demetrius2005, Demetrius2004} (see S1 Appendix for precise mathematical details). This decay rate can be interpreted as a measure of network robustness which, as explained in \cite{Comment_stability}, we use as a synonym of ``stability".

\subsubsection*{Adaptation to the Tangled Nature model}    

\par For each of  the networks associated with TaNa's ecological communities, we calculate the network entropy. To be able to do this we want extend the definition provided above to the case of the TaNa general adjacency matrix, by applying the standard procedure to calculate network entropy after setting $A = Jn$. If all the elements of $A$ were positive,  we would  compute the Markov matrix $P$ associated with $A$ according to Eq \eqref{eq:P_ij}; next, we would calculate its stationary distribution $\boldsymbol{\pi}$; finally, we would derive the network entropy according to Eq \eqref{eq:H}.

\par  However, to compute network entropy, we need to be able to apply the Perron-Frobenius theorem to $Jn$. This, in principle, is not possible, because the effective interaction matrix also contains negative numbers, for all the terms associated with $J_{ij} < 0$. Hence, we focus on the structural aspects of the network, disregarding the specific types of interactions and instead concentrating on their existence. To do this, we consider two modified versions of the effective interaction matrix. We refer to the associated networks as the unsigned and binary versions of the TaNa interaction network \cite{Talaga2023}. 

\par The unsigned network ignores the sign of the interactions. The associated adjacency matrix is denoted $A_\text{\rm uns} = |Jn|$, meaning that each entry is taken as $(A_\text{\rm uns})_{ij} = |(Jn)_{ij}|$. This version does not distinguish between mutualistic, competitive, and predator--prey interactions. A more detailed analysis of the role of the sign of the links is provided later.

\par The binary network ignores both the sign and the strength of the interactions. The entries of the associated adjacency matrix are defined as \( (A_{\text{\rm bin}})_{ij} = 0 \) if \( (Jn)_{ij} = 0 \) and \( (A_{\text{\rm bin}})_{ij} = 1 \) if \( (Jn)_{ij} \neq 0 \).  This consists in a very rough approximation that only highlights the extant species and their connections. 

\par In both cases, network entropy can only be computed if the resulting network is strongly connected. While this condition is necessary to apply the Perron–Frobenius theorem, it does not hold in general. However, in the TaNa model, it is known that if the connectivity $\theta$ is sufficiently high, all species remain part of a single connected component at all times \cite{Anderson2005}. Moreover, because the non-zero interaction strengths are typically drawn from continuous distributions -- as discussed earlier -- it is extremely unlikely for a unidirectional link to exist without a corresponding reverse interaction. This guarantees that the TaNa networks considered in this paper are typically strongly connected, with very few exceptions.

\subsection*{The clustering coefficient}

\par To validate the results obtained for network entropy, we would ideally make use of other stability measures such as loopiness and trophic incoherence \cite{Johnson2017}. However, computing the number of cycles is too numerically demanding and the TaNa interaction networks are not limited to trophic structures (in particular, the frequent absence of basal nodes makes it imposible to compute trophic incoherence).  As a viable alternative, we use the clustering coefficient as a proxy of network loopiness. The global clustering coefficient is defined as the ratio of the number of closed triplets to the total number of all triplets (both open and closed) in the network, equivalent to three times the number of closed triangles divided by the total number of all triplets. The key idea behind this choice is the assumption that directed triangles constitute the predominant type of cycle. While being a rough approximation that most likely only provides a limited description of the real network's structure, some evidence exists that patterns in shorter cycles may be more important than in longer ones, a rationale supported by the fact that the model is fundamentally defined by pairwise interactions. Hence, counting triangles serves as a way to estimate the number of cycles, and consequently relate to the network's loopiness and stability.

\par To calculate the global clustering coefficient, we have employed the dedicated Python's built-in function from the package NetworkX. Details regarding how the coefficient is computed are provided in S1 Appendix.

\section*{Results}

\par To evaluate the temporal change in network entropy as the ecological network associated with the TaNa model evolves, we used an ensemble of 200 simulations of 100,000 generations each. While each stochastic evolution might lead to casual switch between more and less robust networks, analysing an entire ensemble permits to identify possible trends. The parameters used were the following for all simulations: $L = 10, \theta = 0.25, p_{kill} = 0.2, \mu = 1/143, K = 33, N(0) = 1000$. The interaction matrix was generated randomly, with connections drawn from a uniform distribution between -1 and 1. A visual representation of the matrix is provided in S2 Fig. Network entropy was computed on each unsigned and binary network after verifying that they were strongly connected; those few that did not meet this requirement (less than 0.003\%) were simply excluded from the statistics. 

\subsection*{Network entropy and diversity}

\par
\begin{figure}[!h]
\includegraphics[width=\textwidth]{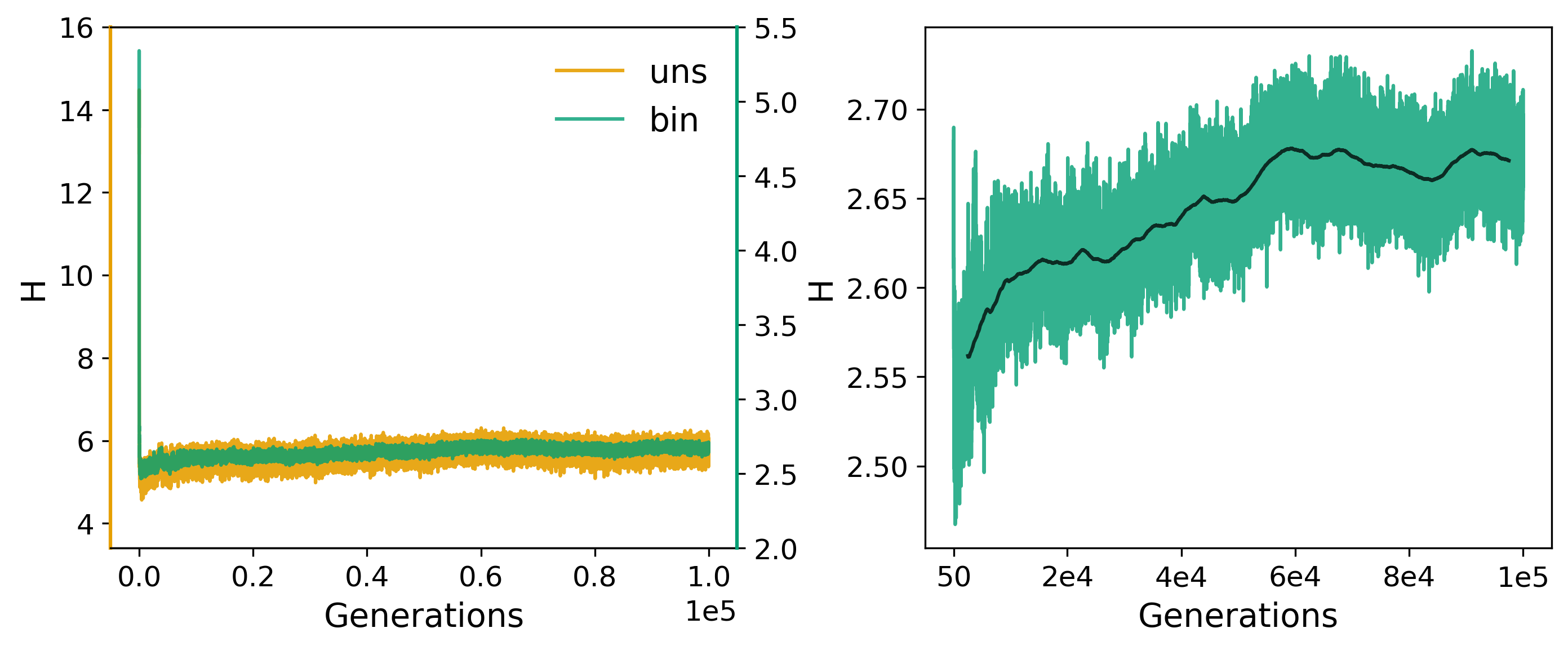}
\caption{{\bf Network entropy in TaNa networks.}
(Left) Curves of the network entropy calculated on the unsigned (orange) and binary networks (green) of the effective interaction matrices representing the evolution of TaNa ecological communities over time. Values are obtained as the mean over an ensemble of 200 simulations of 100,000 generations each (the parameters used appear in the main text). (Right) Zoom on the curve of binary networks in which we neglect the initial plunge, starting from generation 50. The black line represents the average over a 5,000-generation sliding window.}
\label{fig:Network_entropy}
\end{figure}

 The network entropies computed on the binary and unsigned effective interaction matrices, as reported in Fig \ref{fig:Network_entropy}, left panel,  show a closely related behaviour. In addition, after a marked initial drop, both curves show a slow but steady increase over time. 
Interpreting this result in light of the FST suggests that, as the system evolves through successive qESS states under evolutionary pressure, the TaNa ecological communities become increasingly robust. This is consistent with the increase in stability over time predicted by other indicators presented in earlier sections -- like the increase in the duration of qESS states and decrease in the occurrence of transitions -- and provides a computational verification of the FST itself. The evolution towards more stable communities is accompanied by a decrease in the probability to leave an established qESS state. This implies  that the perturbations caused by the emergence of detrimental mutants are reabsorbed faster, making the system persist in the given qESS state for longer. 

\par As highlighted by Fig \ref{fig:Network_entropy}, right panel, the slope of the increase in network entropy decreases with time. This is in line with the fact that the number of transitions per unit time between qESS states decreases as the system ages and that the systemic properties are generated by the search in type space performed during the transitions. On the other hand, the initial drop can be explained by considering that in the first generations the dynamics is affected more by the random initialisation and a high number of transitions. 

\par Indeed, the early-stage ecological networks are generally composed of a higher number of species, with non-selected connections, which makes the network more diverse and hence with a higher entropy when the absolute values of the interactions are considered. However, many of the connections in these young networks are antagonistic, leading to a structure that is, in fact, weaker. Conversely, it will be shown later that as time passes, mutualistic interactions become predominant, making the whole network effectively more robust. 

\par In addition, the transition periods between qESS states typically correspond to higher network entropy, so that in the first generations of the simulation -- where transitions occupy a significant proportion of the time -- the average network entropy can be higher. Conversely, at later stages transitions make up a negligible fraction of the time, and the increase in network entropy must be linked to an effective increase in the robustness of qESS-state networks. 

\par While we have established a clear correlation between increased network entropy and enhanced ecosystem stability, the underlying mechanisms for this relationship are not yet fully appreciated. In fact, a particularly important factor to consider is the temporal variation in network size. As the number of species increases over time, so does the number of possible interactions which, in turn, elevates network entropy. This raises the possibility that, in this context, network entropy may effectively serve as a proxy for species richness.

\par
\begin{figure}[!h]
\includegraphics[width=\textwidth]{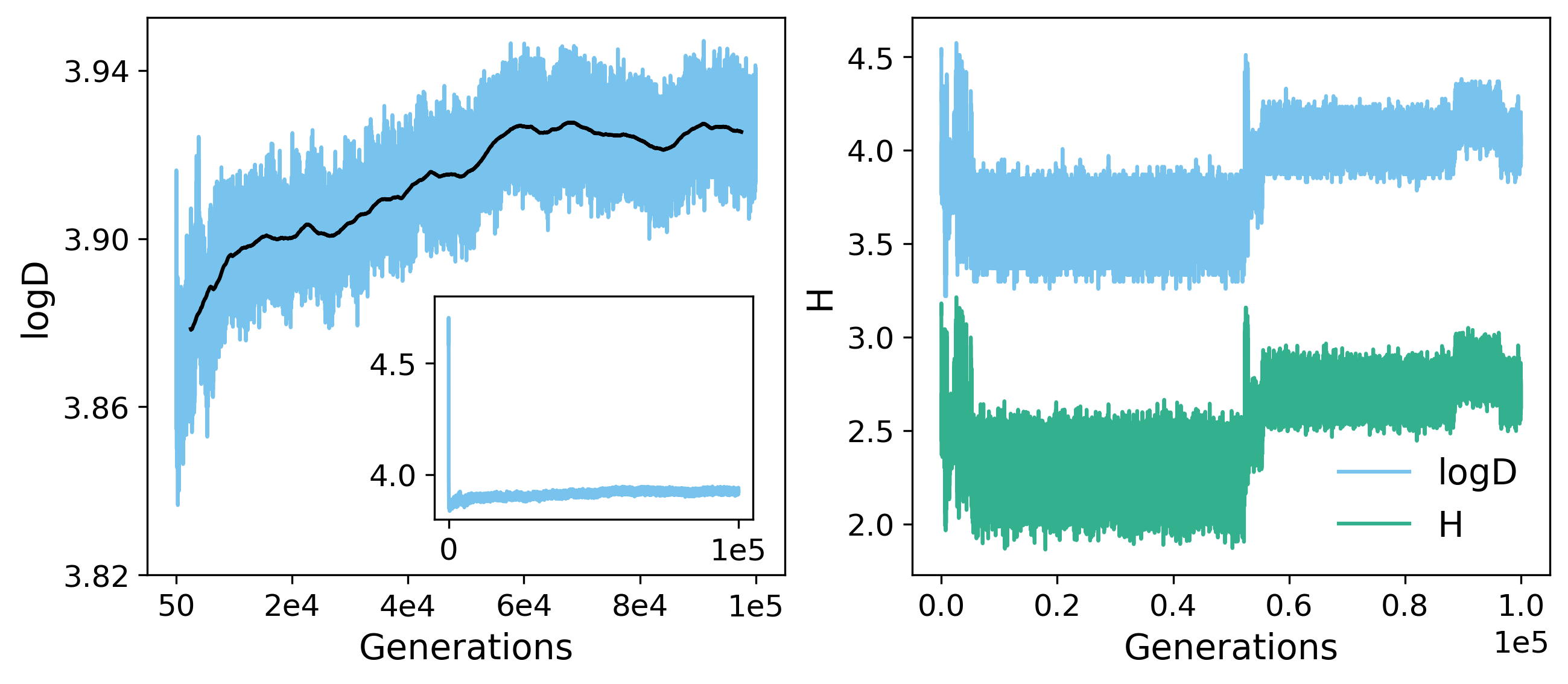}
\caption{{\bf Species diversity in TaNa networks.}
(Left) Time evolution of the logarithm of the average species diversity \cite{Comment_diversity} over the ensemble of 200 simulations. (Right) Plots of the evolution of  network entropy (green) and the logarithm of the diversity (light blue) over time in a selected simulation. The network entropy is computed on the binary version of the networks.}
\label{fig:Diversity}
\end{figure}

 Indeed, our analysis reveals a strong link between network entropy and system diversity \cite{Comment_diversity}. Fig \ref{fig:Diversity}, left panel illustrates the time evolution of the logarithm of the number of distinct species, averaged across an ensemble of simulations. A simple comparison with Fig \ref{fig:Network_entropy} clearly shows that the trajectory of species diversity closely mirrors that of network entropy. This correspondence is further elucidated in Fig \ref{fig:Diversity}, right panel, where we compare the evolution of network entropy with the logarithm of species diversity for a selected simulation. The ratio between these two quantities exhibits only minor oscillations around a constant value -- a behaviour that is consistently observed in all simulations.

\par Notably, while network entropy closely tracks species diversity, it does not reach the theoretical maximum, which would correspond to the logarithm of the number of species. The reason for this deviation is that the networks considered are not fully connected. Consequently, the measured entropy remains lower than the upper bound. This observation implies that the number of effective connections in the TaNa model is directly influenced by the species richness of the system. More broadly, it underscores the conclusion that ecosystem stability is intimately connected to, and perhaps governed by, its biological diversity.

\subsection*{The clustering coefficient}

\par
\begin{figure}[!h]
\includegraphics[width=1\textwidth]{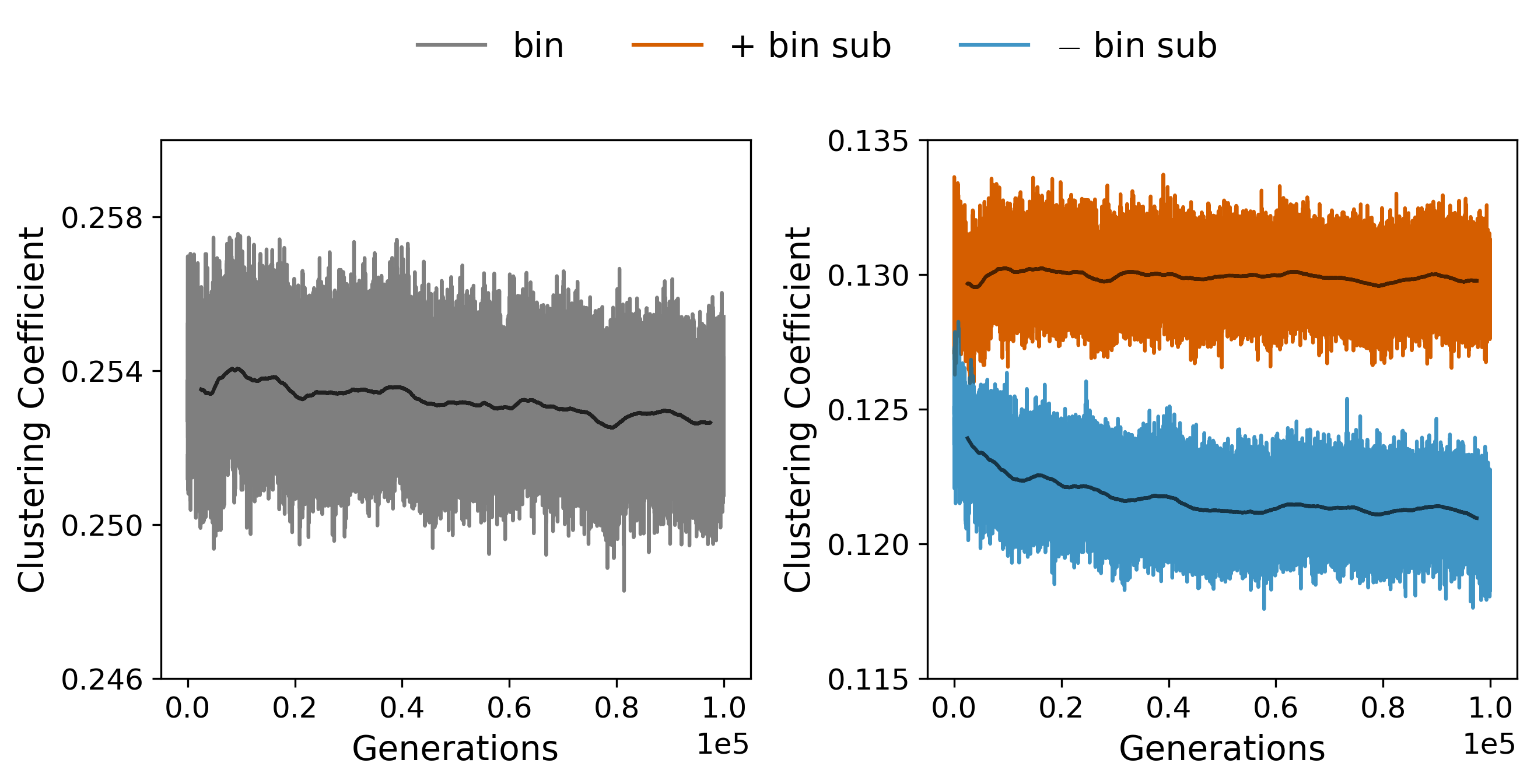}
\caption{{\bf Clustering in binary TaNa networks.}
(Left) Plot of the global clustering coefficient calculated on the binary networks of the effective interaction matrices that represent the evolution of TaNa ecological communities over time. (Right) Same quantity calculated on the binary subnetworks of the positive (red) and negative (blue) components of the effective interaction matrices. In both panels, the values are obtained as the mean on the same ensemble of 200 simulations of 100,000 generations, while the black lines represent the average over a 5,000-generation sliding window.}
\label{fig:Clustering_binary}
\end{figure}

 Further information about the evolution of the system can be deduced by studying the clustering coefficient, according to the methodology illustrated earlier. Given the similarity between the unsigned and the binary  version of the TaNa networks, we just focus on the latter. The results for the ensemble of 200 simulations  are reported in Fig \ref{fig:Clustering_binary}, left panel. The global clustering shows, on average, a slow but steady decrease over time. This, if clustering is a proxy of network stability, is in contrast with previous results. In fact, since we know that the robustness of TaNa ecological communities increases over time, making them more connected, we should expect an increase in the average clustering, as long as it serves as a proxy of network robustness.

\par However, we recall that the binary transformation of the original adjacency matrix $Jn$ describing TaNa ecological communities is a rough approximation. In fact, network connections cannot be considered all equal to each other, because crucial differences distinguish positive interactions from negative ones. While the former enhance the reproduction capability of the receiving node, the latter inhibit it. We therefore need to take into account such differences.

\par To do this, we propose a simple method to study the two types of interactions separately. We divide the interaction network described by $Jn$ into its positive and negative components, $Jn^+$ and $Jn^-$, and thereafter transform each of these subnetworks into binary networks $A_{\text{bin}}^+$  and $A_{\text{bin}}^-$ by imposing $(A_{bin}^\pm)_{ij} = 0$ if $(Jn^\pm)_{ij} = 0$ and $(A_{\text{bin}}^\pm)_{ij} = 1$ if $(Jn^\pm)_{ij} \neq 0$. In this way, $A_{\text{bin}}^+$ describes the pure structure of the positively interacting subnetwork, while $A_{\text{bin}}^-$ describes the structure of the negatively interacting one.

\par We proceed by calculating the global clustering on each subnetwork. The results are reported in Fig \ref{fig:Clustering_binary}, right panel. It is evident that, over time, the positive subnetwork is characterised by a constant clustering coefficient, implying that its robustness is maintained. Conversely, the negative subnetwork is affected by a decrease in the clustering coefficient, meaning that its robustness weakens accordingly. As the stability of the negative subnetwork decreases while that of the positive one remains constant, the overall stability of the whole network will increase. Notice also that the clustering of the negative subnetwork remains lower than that of the positive one at all times.

\subsection*{Details of the importance of the two subnetworks}

\par We have then studied further the ecological meaning of the two subnetworks consisting, respectively, of the subset of positive  links only and negative  links only in the matrix $Jn$.  
In particular, we studied the time evolution of the distribution of the entries of the interaction matrix $Jn$, grouping them in terms of core and cloud. The core was chosen as the set of types whose occupancy exceeds 5\% of the most populous type, while the remaining types constitute the cloud \cite{Becker2014}. Next, we distinguish the interactions on the basis of the character of the two nodes they connect: core-on-core, cloud-on-cloud, core-on-cloud, and cloud-on-core.

\par
\begin{figure}[!h]
\includegraphics[width=\textwidth]{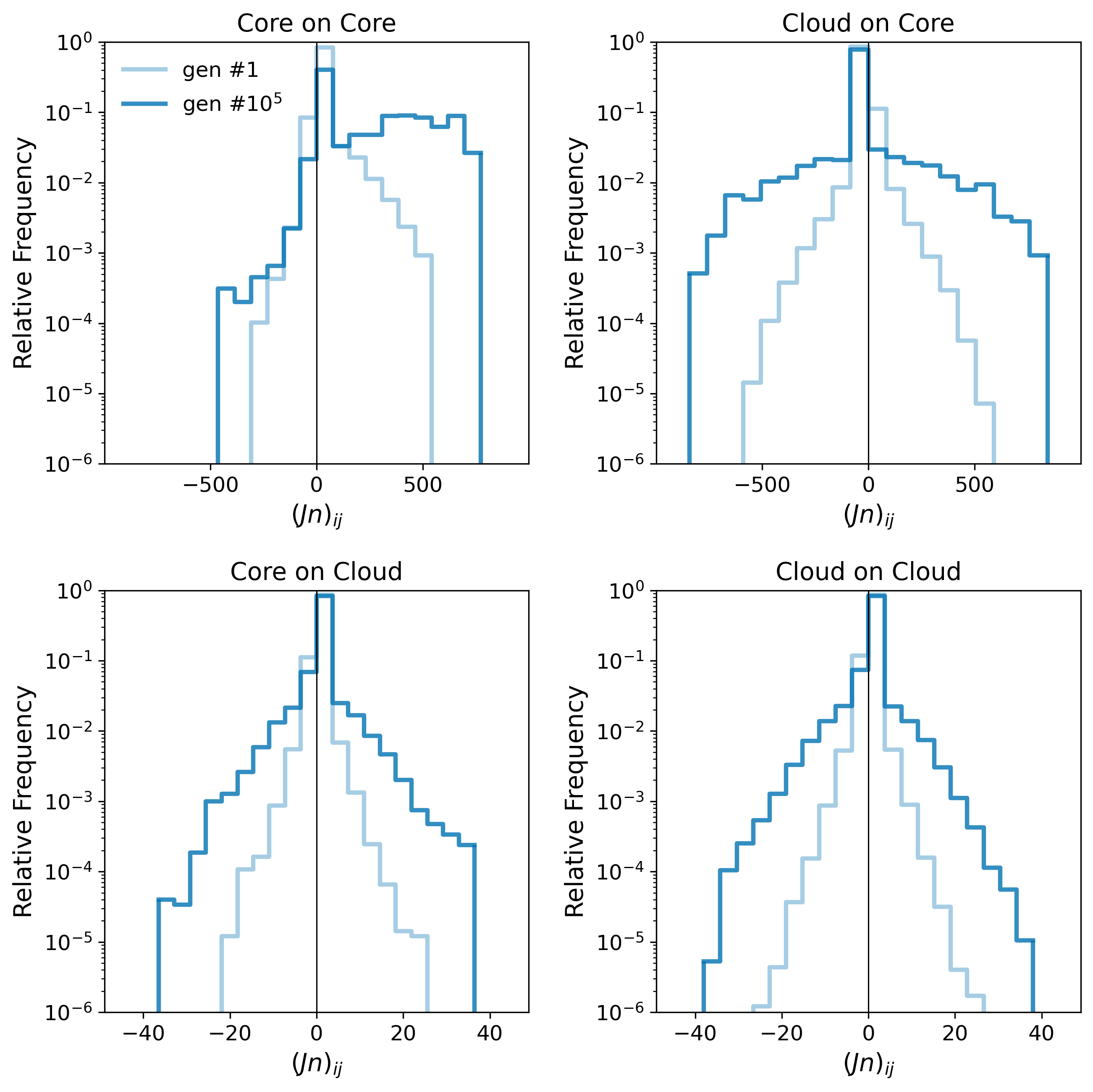}
\caption{{\bf Effective interaction of the core and the cloud.}
Plots of the distributions of the effective directed interactions as defined by the entries of $Jn$, grouped depending on the belonging to the core or to the cloud of the edges connected. The distributions are obtained by  averaging over the ensemble of 200 simulations. The x axes report the strength of the effective interaction, while the y axes indicate the relative frequency of histogram bins. Each plot reports the curve obtained after one generation and the one obtained after 10\textsuperscript{5} generations. The bins are symmetric around the zero.}
\label{fig:Interaction_core_cloud}
\end{figure}

 The results for the ensemble of 200 simulations are reported in Fig \ref{fig:Interaction_core_cloud}. After one generation, when the networks have not yet matured but have already started to move away from the completely random distributions assigned as initial conditions, the distributions involving cloud elements appear approximately symmetric, but with some crucial deviations. Namely, cloud species tend to establish slightly positive interactions between each other and have a slightly negative influence on the core, while the core elements show a slightly positive influence on the cloud. Conversely, the core-on-core distribution displays a hint of positive shoulder, indicating that core elements are generally connected by positive links. 

\par The following interpretation can be drawn for the distributions shown in Fig \ref{fig:Interaction_core_cloud}. Core species host the largest portion of the population in the ecological community. Core types sustain each other through positive (i.e.  mutualistic) interactions. Cloud types, on the other hand, exhibit  weak mutualistic links with other cloud types, allowing these to exist for some time but not enough to reach any dominance in the system. In addition, cloud types are typically coupled to core types through weak negative (i.e. antagonistic or parasitic) couplings. The prominent presence of core types in the system allows for the survival of the cloud, while the substantial weight of the core allows it to withstand, at least for a while,  the detrimental effects of cloud types. 

\par After 2,000 generations, the situation has moderately evolved. The distributions of couplings involving the cloud appear to become broader, but remain almost symmetric apart from the region around zero, where their skewness towards slightly negative (cloud-on-core) or slightly positive values (core-on-cloud, cloud-on-cloud) is accentuated (notice the log scale). This suggests that the general cloud-cloud and cloud-core relations described earlier become established and gain strength. On the other hand, the core-on-core distribution develops a more pronounced positive shoulder, exhibiting a second broad peak at very high values of $(Jn)_{ij}$ (around 500) alongside the usual peak located just to the right of zero. This peak is still dominant but less prominent. 

\par As the system evolves, core types tend to become strongly mutualistic, creating a solid structure that sustains the entire ecological community. Conversely, cloud types continue to have a mildly antagonistic relation to the  core, while not supporting each other enough to become predominant. Together with the results for the clustering, these findings suggest that, over time, positive interactions within the core develop to become the backbone of the entire network, while the negative effect of the cloud on the core may be reduced.
\section*{Discussion}

\par Analogously to classical thermodynamic systems, the collective behaviour of evolutionary ecological systems emerges from the interplay of numerous microscopic interactions. However, extending thermodynamic methods to capture their directionality is not straightforwardly feasible, as living systems operate out of equilibrium. In this work, we have evaluated alternative approaches based on a network formulation of ecological communities. The use of the agent-based TaNa model allows one to account for the main characteristics of empirical ecological networks while keeping a simple formulation. 

\par Crucially, the complex adaptive macroscopic behaviour of the TaNa emerges solely from the underlying microscopic dynamical rules. In this model, the combination of mutation, interaction, and death gives rise, when the interaction parameters are appropriately tuned, to an intermittent behaviour at the macroscopic level, in which ecological communities abruptly replace each other evolving towards increasingly more stable configurations. This stability can be deduced from the increase in the duration of the metastable qESS states and the reduction in the occurrence of transitions.

\par Defining a precise analytic measure that could capture this directional trend was the main goal of the present work. In particular, rewriting the TaNa as a network problem has permitted to test appropriate stability measures capable of revealing a temporal directionality trend that follows ecological evolution. The translation of occupancy data for TaNa ecological communities into proper ecological networks is not without pitfalls. However, some intuitive simplifications have allowed us to grasp some basic evolutionary properties. 

\par Specifically, focusing on the topological structure of the   strongly connected networks, i.e, neglecting the sign of the interactions, has allowed us to apply a stability measure known as network entropy. The motivation for this approach stems from a well-established mathematical framework, encapsulated by the FST, which links network entropy to the system's rate of recovery following perturbations\cite{Demetrius2013}. Hypothesising that the mechanisms responsible for enhanced stability may be encoded in the complex architecture of interconnections among types, network entropy served as a natural metric in that it quantifies the number of distinct pathways that the network allows. Indeed, simulations show that the average network entropy increases over time.

\par However, network entropy might not directly capture the core drivers of ecological dynamics. In fact, a critical aspect of the formulation adopted is that the very network size fluctuates over time, reflecting the dynamic number of extant species. Since the number of distinct paths in a network is inherently related to its size, we hypothesised that these size fluctuations could be the primary driver of changes in network entropy. Led by this, we have verified that the diversity of ecological communities, understood here as species richness, closely tracks the behaviour of network entropy. This suggests that the number of connections in the TaNa is directly coupled to the number of species in the system and reveals that the stability of TaNa ecosystems is tightly linked to its diversity.

\par This finding is not novel within the field of ecology. Some twenty years ago, Hooper and colleagues had already hypothesised, based on a vast literature review, that species richness may affect the stability of ecosystem properties \cite{Hooper2005}. In their work, the authors point out that the stability of an ecosystem is affected by a variety of factors, the most important of which is likely to be the functional diversity of the species that constitute it. In fact, it is only as long as an increase in the number of species translates into an increase in species functional diversity that the species richness determines the robustness of ecosystems to perturbations \cite{Hooper2005}. Encouragingly, some experimental findings seem to have confirmed this connection in the last couple of decades; see e.g. \cite{Biswas2011}. Indeed, both computational and empirical studies -- see \cite{Vila2019} and references therein -- have pointed out that genotypic richness enhances the robustness of ecosystems to invasion. In the current work, we have contributed to reinforce this hypothesis by proposing a verification based on a computational model, the TaNa, that exhibits an increasingly stable behaviour with diversity, and on the exploitation of a mathematical measure, network entropy, which is proven to predict the robustness of networks to perturbations through the FST. 

\par Our treatment relies on a few somewhat strict assumptions about the nature of the networks considered, limiting its scope to a specific set of cases. Nonetheless, it sheds light on some important dynamics at the core of ecosystems' functioning, proposing that stability and diversity are intimately connected through evolutionary mechanisms and establishing a connection between well-established mathematical findings and field work.

\par Further analysis has focused on more detailed structural features of the interaction network, partly to address the limitations introduced by the earlier simplifying assumptions. In particular, we have examined the temporal evolution of the clustering coefficient within the subnetworks defined by positive and negative interactions, as well as the evolution of the interaction strengths and signs between core and cloud species. In this context, the clustering coefficient was interpreted as a proxy for the prevalence of cycles in the network, which are in turn linked to network stability.

\par The results indicate that the subnetwork of negatively interacting species tends, on average, to weaken over time, thus contributing to an overall increase in the system's stability. Simultaneously, the network of dominant species -- those composing the ecological core -- evolves towards increasingly entangled mutualistic configurations, which are more resistant to destabilisation by antagonistic (mutant) invaders -- a trend that is consistent with certain experimental observations \cite{Leigh2010, Venkataram2023}.

\par As before, the way we delineate network clusters -- formed on positive versus negative interactions, or core versus cloud species -- remains somewhat arbitrary. Although these partitions capture important structural aspects of the system, they inevitably lead to a loss of information about connections that span across cluster boundaries -- for instance, the interactions \textit{between} the positively and negatively interacting subnetworks. It is also possible that the most dynamically relevant groupings differ from those we have imposed. In particular, the links connecting different clusters may follow more intricate structural patterns and carry critical information about the evolution and stability of the system. 

\par Taken together, our results suggest that, although a general quantitative measure for the directionality of evolutionary dynamics remains elusive, certain macroscopic observables -- most notably species diversity and the degree of mutualism -- emerge as promising candidates to characterise long-term ecological trends. Among these, diversity stands out for its broad applicability: it is a conceptually simple quantity that can be readily measured across virtually any ecological system, unlike more delicate measures such as network entropy. After all, experimental findings increasingly support the idea that diversity and stability, as well as stability and mutualism, are intimately connected through adaptive evolutionary mechanisms \cite{Vila2019, Leigh2010, Venkataram2023}.

\par Building on our results, we suggest that future work should focus on validating the proposed link between diversity and evolutionary trends, with the aim of assessing its predictive power in broader ecological settings. In particular, it would be valuable to explore whether this connection holds in more general contexts -- such as in networks that are not strongly connected -- to determine the solidity and applicability of diversity as a universal indicator. Furthermore, alternative ways of partitioning the network could be explored, allowing systematic analysis of structural properties using methods inspired by Structural Balance Theory \cite{Talaga2023}. Finally, we propose that key quantities identified in this study -- namely network entropy, species diversity, and the clustering coefficients of the positive and negative interaction subnetworks -- be tested on empirical ecological networks, following approaches similar to those developed in \cite{PascualGarcia2020}.

\section*{Supporting information}

\par {\bf S1 Appendix. Mathematical details.}

\par {\bf S2 Figure. Interaction matrix for the simulation ensemble.}
(Main) Heatmap of the interaction matrix used for all the simulations in the present study. Along the axes are the type identifiers, reported in arbitrary order, while the colourbar indicates the interaction magnitude. The matrix entries are randomly drawn from a uniform distribution in [-1,1], after the interacting pairs are selected with a probability given by the connectance $\theta = 0.25$. (Inset) Zoom on the interactions between the first 10 types. It is evident that, when disregarding the interaction sign and strength, the matrix is essentially symmetric.

\section*{Acknowledgments}
AM is grateful for the hospitality of the Centre for Complexity Science at Imperial College London, made possible through the Erasmus+ exchange scheme with the Department of Physics at Imperial. Additional support was supplied by the Galilean School of Higher Education, Padua, and the University of Padua. AM thanks the members of the Centre for inspiring discussions and extends a special thanks to Hardik Rajpal for providing his data and assisting with Imperial’s High-Performance Computing cluster. The authors also thank Lloyd Demetrius for fruitful discussions.

\nolinenumbers

%
%
%

\end{document}